\documentclass[
]{ceurart}

\usepackage{listings}
\usepackage{algorithm}
\usepackage{algpseudocode}
\usepackage{caption}
\usepackage[justification=centering]{caption}
\usepackage{subcaption}
\usepackage{censor}
\begin{document}

\copyrightyear{2025}
\copyrightclause{Copyright for this paper by its authors.
  Use permitted under Creative Commons License Attribution 4.0
  International (CC BY 4.0).}

\conference{ECOM'25: SIGIR Workshop on eCommerce, Jul 17, 2025, Padua, Italy}

\title{Improving Ad matching via Cluster-Adaptive Keyword Expansion and Relevance tuning}


\author{Dipanwita Saha}[%
email=dipsaha@ebay.com,
]
\address{eBay Inc., San Jose, CA, USA}

\author{Anis Zaman}[%
email=anzaman@ebay.com,
]

\author{Hua Zou}[%
email=huazou@ebay.com,
]

\author{Ning Chen}[%
email=ningchen@ebay.com,
]

\author{Xinxin Shu}[%
email=xshu@ebay.com,
]

\author{Nadia Vase}[%
email=nvase@ebay.com,
]

\author{Abraham Bagherjeiran}[%
email=abagherjeiran@ebay.com,
]

\begin{abstract}
In search advertising, keyword matching connects user queries with relevant ads. While token-based matching increases ad coverage, it can reduce relevance due to overly permissive semantic expansion. This work extends keyword reach through document-side semantic keyword expansion, using a language model to broaden token-level matching without altering queries. We propose a solution using a pre-trained siamese model to generate dense vector representations of ad keywords and identify semantically related variants through nearest neighbor search. To maintain precision, we introduce a cluster-based thresholding mechanism that adjusts similarity cutoffs based on local semantic density. Each expanded keyword maps to a group of seller-listed items, which may only partially align with the original intent. To ensure relevance, we enhance the downstream relevance model by adapting it to the expanded keyword space using an incremental learning strategy with a lightweight decision tree ensemble. This system improves both relevance and click-through rate (CTR), offering a scalable, low-latency solution adaptable to evolving query behavior and advertising inventory.
\end{abstract}


\maketitle

\section{Introduction}
In online advertising, traditional keyword-based systems rely on exact token overlap, limiting matches to queries sharing lexical tokens with advertiser-specified keywords. For example, a keyword like “iPhone case” would miss semantically related queries such as “Apple phone cover,” reducing ad reach and limiting advertisers who cannot anticipate all phrasing variations.
We introduce a document-side semantic keyword expansion system, enriching seller-provided keywords without altering user queries. Using a pre-trained siamese neural network, we generate dense embeddings for keywords and perform nearest neighbor searches to find semantically similar terms. This allows the keyword “iPhone case” to expand to terms like “Apple phone case” or “smartphone cover for iOS,” enabling broader, relevant matches without requiring exact token overlap. To control precision, we implement a cluster-based adaptive thresholding strategy. We segment embeddings using k-means clustering, setting local similarity thresholds based on each cluster’s semantic density. Smaller, denser clusters receive stricter thresholds to avoid irrelevant matches, while larger clusters representing broader, ambiguous concepts have more permissive thresholds to improve recall. This ensures high-quality matching tailored to semantic context.
Since ads are ultimately displayed as specific items (products), expansions must closely align with the actual products they represent. Poorly matched expanded keywords could result in irrelevant ads being shown, decreasing click-through rates (CTR). To address this, we enhance the existing relevance model(Gaussian regression) with an incrementally trained lightweight decision tree ensemble, leveraging human-labeled relevance data specifically for expanded matches. This targeted approach ensures effective quality control, enabling accurate relevance assessment of new matches while maintaining system stability. Collectively, these components form a robust, scalable pipeline capable of daily refreshes, significantly enhancing query coverage and match quality at production scale.

\section{Related Work}

Semantic keyword expansion is widely explored in advertising, information retrieval, and content generation to improve keyword coverage and ad relevance. Early work by Azimi et al.~\cite{adsrewrite2015} leveraged external search engine results to rewrite rare ad keywords into more common forms, improving matching without altering user intent. Similarly, Mandal et al.~\cite{Mandal2019QueryRU} introduced automated synonym extraction to rewrite queries for enhanced relevance in e-commerce searches.
In recent years, embedding-based methods have become prevalent. Grbovic et al.~\cite{10.1145/2911451.2911538} demonstrated how semantic embeddings of user queries and ads significantly improve ad matching performance in sponsored search. Mandal et al.~\cite{mandal2023semanticequivalenceecommercequeries} further developed siamese models specifically trained for capturing semantic equivalence between e-commerce queries, enhancing the quality of retrieval by recognizing intent equivalence. In large-scale e-commerce settings, Li et al.~\cite{Li2022QueryRI} presented a comprehensive query rewriting approach employed in Taobao search to bridge vocabulary gaps between queries and products, substantially boosting user experience and retrieval effectiveness. Similarly, Wu et al.~\cite{chen2018e2e} proposed an end-to-end neural matching framework integrating vector-based retrieval and neural ranking, demonstrating significant improvements in ad matching for e-commerce.Generative models also emerged as a powerful tool for keyword augmentation. Shi et al.~\cite{shi-etal-2021-keyword} employed generative sequence-to-sequence methods combined with trie-based search for effective keyword augmentation, facilitating richer product descriptions and improved ad targeting.
Our work distinguishes itself by performing document-side keyword expansion using pre-trained siamese embeddings~\cite{mandal2023semanticequivalenceecommercequeries} , specifically designed to maintain compatibility with token-based ad retrieval systems. We further enhance semantic matching through adaptive cluster-specific similarity thresholds, which manage precision and recall effectively in diverse semantic regions, akin to the density-aware filtering. Finally, our relevance filtering employs gradient boosting decision trees (GBDT), an approach noted for efficiency and performance in ad ranking scenarios. Specifically, we adopt the GBM method developed by Bischl et al.~\cite{bischl2016mlr}, leveraging its efficiency and incremental learning capabilities to maintain system stability while dynamically adapting to evolving data.

\section{Proposed solution}

This section details our approach to expanding seller keyword capabilities through semantic similarity. The proposed system, identifies keyword variants with similar meanings using embedding-based nearest neighbor search. Our methodology combines advanced embedding techniques with clustering-based thresholding and relevance model adjustments to optimize both coverage and precision.

\subsection{Semantic Expansion Framework}

\subsubsection{Expansion set generation}
\label{sec:emb_generation}

For semantic representation of keywords, we leveraged a pre-trained embedding model developed by Mandal et al.~\cite{mandal2023semanticequivalenceecommercequeries}, originally designed for modeling semantic equivalence between e-commerce queries. This model was trained using a two-tower architecture with contrastive learning, incorporating a micro-BERT encoder fine-tuned on eBay titles and a query-category classifier to enhance semantic representations. In our setting, because keywords are typically subsets of buyer queries, this model is particularly well-suited to generate embeddings that reflect partial query semantics. Rather than training a new model from scratch, we reused this optimized siamese network to efficiently generate high-dimensional vector embeddings for AdKeywords. The model encodes each keyword into a dense semantic space where similar concepts appear close together. This enables robust identification of semantically related keyword variants, even when surface forms differ significantly. We measure similarity using cosine distance between vectors.
Using the generated embeddings, we implemented a nearest neighbor search to identify semantically similar keywords. We employed the FAISS \cite{johnson2019billion} library with a flat index structure to ensure accuracy.
\subsection{Cluster-based Thresholding}
While our batched nearest neighbor search strategy ensures scalability and maintains recall, it introduces a new challenge—controlling the precision of the expanded keyword set. Not all nearest neighbors retrieved via embedding similarity are equally relevant, particularly in regions of the embedding space where semantically diverse keywords are densely packed. A uniform similarity threshold applied globally fails to account for these variations in semantic density, leading to inconsistent match quality: overly strict in sparse regions and overly permissive in dense ones. To address this, we introduce a cluster-based thresholding mechanism that adapts the expansion sensitivity based on the local structure of the embedding space.







We partitioned the keyword embedding space into $M$ clusters namely, $\mathcal{C} = \{C_1, C_2, ..., C_M\}$ using $k$-means clustering where each cluster $C_m$ has a centroid $\mu_m$.
For each cluster $C_m$, we computed the distribution of distances between each point and the cluster centroid $D_m = \{sim(k, \mu_m) | k \in C_m\}$. We determined a distance threshold $\tau_m$ as the $p$-th quantile of the distance distribution that optimizes our internal relevance metric across all clusters:

\begin{equation}
\tau_m = \text{Quantile}(D_m, p)
\end{equation}

The quantile $p$ is chosen as a global value and is constant for all clusters to maximize impression while maintaining quality. This approach allows for varying thresholds across different semantic clusters, accommodating the heterogeneous nature of keyword relationships. For any new keyword $k$ assigned to cluster $C_m$, we only consider semantic variants $v$ where $sim(k, v) \leq \tau_m$.


We determined a distance threshold $\tau_m$ for each cluster as the $p$-th quantile of the intra-cluster similarity distribution. The value of $p$ is treated as a global hyperparameter that controls the precision-recall tradeoff across all clusters. A higher $p$ results in more permissive thresholds, allowing broader keyword expansions. Interestingly, we observed a positive correlation between cluster size and threshold—larger clusters, often corresponding to broader or more ambiguous e-commerce concepts, receive higher thresholds to preserve recall, while smaller, denser clusters are assigned stricter thresholds to maintain precision. To select the optimal value of $p$, we conducted extensive experiments evaluating the impact of different quantile settings on true positive rate (TPR) and impression lift. Details of this evaluation, including TPR trends across quantiles and the effect of post-processing filters, are provided in Section~\ref{sec:exp_results}.

\subsection{Relevance Model Adjustment}

Cluster-based thresholding filters weak keyword expansions but doesn't directly ensure expanded keywords align with campaign items. Advertisers associate keywords with products; during retrieval, a keyword match retrieves these items. However, expanded keywords may only weakly relate to associated products, causing irrelevant matches and lowering user experience. For instance, expanding “Apple Watch accessories” to “smartwatch charger” could incorrectly trigger Apple Watch bands for Samsung charger queries.

Our proposed solution builds upon an existing Gradient Boosted Decision Tree (GBDT) model trained with more than 100 trees on a large dataset of labeled human judgments. The base relevance model is a Gaussian regression model trained in a pointwise manner where human judgment labels, based on a five-point scale(Perfect, Excellent, Good, Fair, and Bad), are used as training targets. To incorporate new sponsored listings while maintaining model stability, we adopt an incremental learning strategy by training a small number of additional trees on a small dataset containing human-judged relevance scores for these new listings. Let \( f_{\text{GBDT}}(x) \) represent the baseline GBDT model with \( T \) trees. To refine relevance predictions, we introduce \( t_r \) additional shallow trees (\( t_r = < 2 \)) of depth $<= 5$, which adjust the baseline model’s outputs. The updated model is refereed as stacked model and is given by:
\begin{equation}
f_{\text{adj}}(x) = f_{\text{GBDT}}(x) + \sum_{i=1}^{t_r} g_i(x)
\end{equation}
where \( g_i(x) \) are the additional trees trained on a small dataset of new sponsored listings. The training objective is to minimize the error between model predictions and human-judged relevance scores. Given a dataset \( D = \{(x_j, y_j)\}_{j=1}^{N} \), where \( y_j \) represents human-judged relevance scores, the optimization is performed by minimizing the loss $\mathcal{L} = \sum_{j=1}^{N} \ell \left( f_{\text{adj}}(x_j), y_j \right)$ where \( \ell(\cdot) \) is a suitable loss function, such as mean squared error (although Huber loss is an alternative as well). 
Performance evaluation is conducted by comparing the adjusted model’s predictions against human-judged scores on a holdout set consisting of listing from keyword expansion module, ensuring improved alignment with human judgments while preventing overfitting. Going forward, there are two approaches to revising the relevance model: (1) retraining the entire model while incorporating a substantial number of examples from semantic keyword expansion, which can be time-consuming; or (2) more immediately, automatically retraining the additional trees on new inventory at regular intervals to adapt to evolving listings.

\textit{Market specific relevance threshold tuning:} Items introduced through keyword expansion exhibited a different predicted relevance distribution compared to those from other ad retrieval methods. As a result, we adjusted the relevance filtering threshold for these items. To determine the threshold, $t_{rel}$, we conducted offline relevance filter simulations using a separate holdout dataset containing human judgments for both keyword expansion items and other recalled items.
The effectiveness of this approach depends on several factors. It is most effective when the baseline GBDT model captures relevant patterns but shows biases or domain shifts, allowing additional trees to refine predictions. This method is beneficial when new data is limited, as retraining from scratch risks overfitting. However, it assumes residual errors are structured enough for meaningful adjustments, and if the baseline model has already extracted key features, gains from stacking may be limited. Finally, careful tuning of the loss function and hyperparameters is essential to prevent overfitting and ensure stability.

\subsection{System Implementation}

We implemented the semantic expansion system as a combination of a service and an Airflow pipeline that supports daily updates. When a new campaign is created, any newly introduced keyword is first embedded and assigned to the nearest semantic cluster. Based on this assignment, a cluster-specific similarity threshold is retrieved. The system then performs a nearest neighbor search in the embedding space to identify semantically similar candidate keywords. These candidates are filtered using the cluster threshold. The resulting expanded keywords are matched to buyer queries, and the associated items in the campaign are considered for ad serving. To ensure only high-quality matches are retained, a specialized decision tree adjusts the relevance score, and a market-specific threshold is applied to finalize the match. This implementation allows for scalable processing of new keywords while maintaining consistent quality standards across the system. Algorithm~\ref{alg:broadmatch} shows the end-to-end pipeline implemented at run-time.

\begin{figure}[t]
  \centering
  \includegraphics[width=\linewidth]{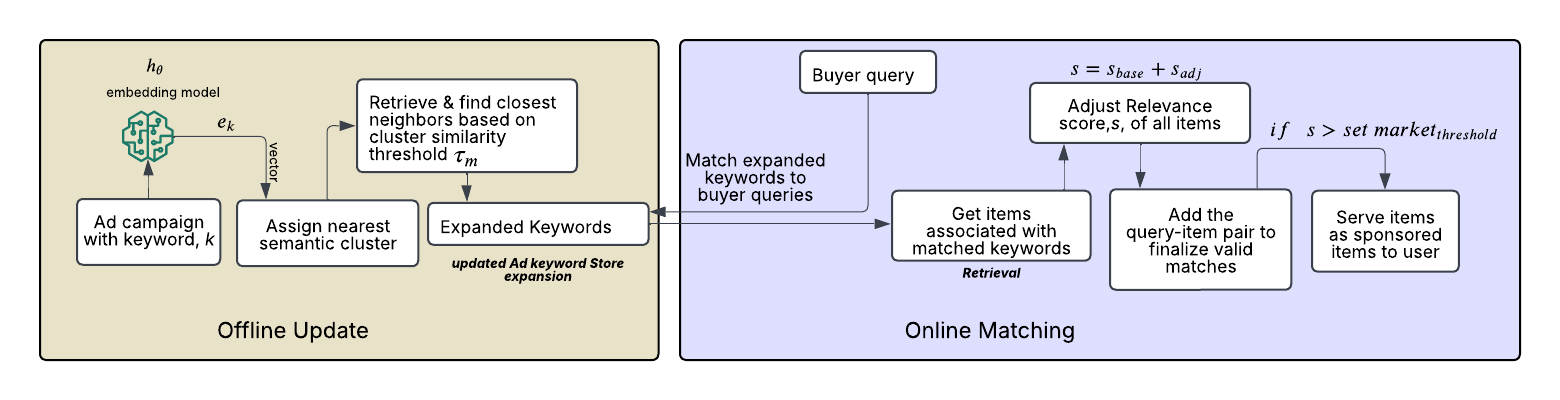}
  \caption{Flowchart of the offline and online process}
  \label{fig:elbow_wcss}
\end{figure}

\begin{algorithm}[t]
\caption{Keyword Expansion and Item-Level Relevance Filtering}
\label{alg:broadmatch}
\footnotesize
\begin{algorithmic}[1]  
\Require New keywords $\mathcal{K}_{\text{new}}$, embedding model $h_{\theta}$, cluster centroids $\{\mu_j\}$, thresholds $\{\tau_j\}$, FAISS index, base relevance model $f_{\text{base}}$, adjustment model $f_{\text{adj}}$, market thresholds $\{T_m\}$
\Ensure Valid query-item matches $\mathcal{M}$
\State $\mathcal{E} \gets \emptyset$, $\mathcal{M} \gets \emptyset$
\For{each keyword $k \in \mathcal{K}_{\text{new}}$}
    \State $\mathbf{e}_k \gets h_{\theta}(k)$
    \State $C_m \gets \arg\min_{j} d(\mathbf{e}_k, \mu_j)$
    \State $N_k \gets \text{FAISS.search}(\mathbf{e}_k)$
    \State $V_k \gets \emptyset$
    \For{each $n \in N_k$}
        \If{$\text{sim}(\mathbf{e}_k, \mathbf{e}_n) \geq \tau_m$}
            \State $V_k \gets V_k \cup \{n\}$
        \EndIf
    \EndFor
    \State $\mathcal{E}[k] \gets V_k$
\EndFor

\For{each $(k, V_k) \in \mathcal{E}$}
    \For{each $n \in V_k$}
        \State Match query $q$ to $n$
        \State Retrieve items $I_n$
        \For{each item $i \in I_n$}
            \State $s_{\text{base}} \gets f_{\text{base}}(q, i)$
            \State $s_{\text{adj}} \gets f_{\text{adj}}(q, i)$
            \State $s \gets s_{\text{base}} + s_{\text{adj}}$
            \State $T \gets T_{\text{Market}(i)}$
            \If{$s \geq T$}
                \State $\mathcal{M} \gets \mathcal{M} \cup \{(q, i)\}$
            \EndIf
        \EndFor
    \EndFor
\EndFor
\State \textbf{return} $\mathcal{M}$
\end{algorithmic}
\end{algorithm}

\section{Experimental Results}
\label{sec:exp_results}
\subsection{Clustering and thresholding}
We used k-means clustering separately for each market (e.g., US, UK, AU) to preserve regional semantic nuances. For instance, “coach” may refer to a luxury brand in the US but a sports trainer in the UK or AU. We employed k-fold resampling to ensure clustering stability, holding out subsets in each fold and evaluating assignment consistency and intra-cluster compactness. The optimal number of clusters (1000) was chosen using the elbow method on the average within-cluster sum of squares (WCSS) across folds. Given steady corpus growth, we plan periodic cluster retraining to maintain semantic precision and mitigate drift.

\begin{figure}[t]
  \centering
  \begin{subfigure}[b]{0.48\linewidth}
    \centering
    \includegraphics[width=\linewidth]{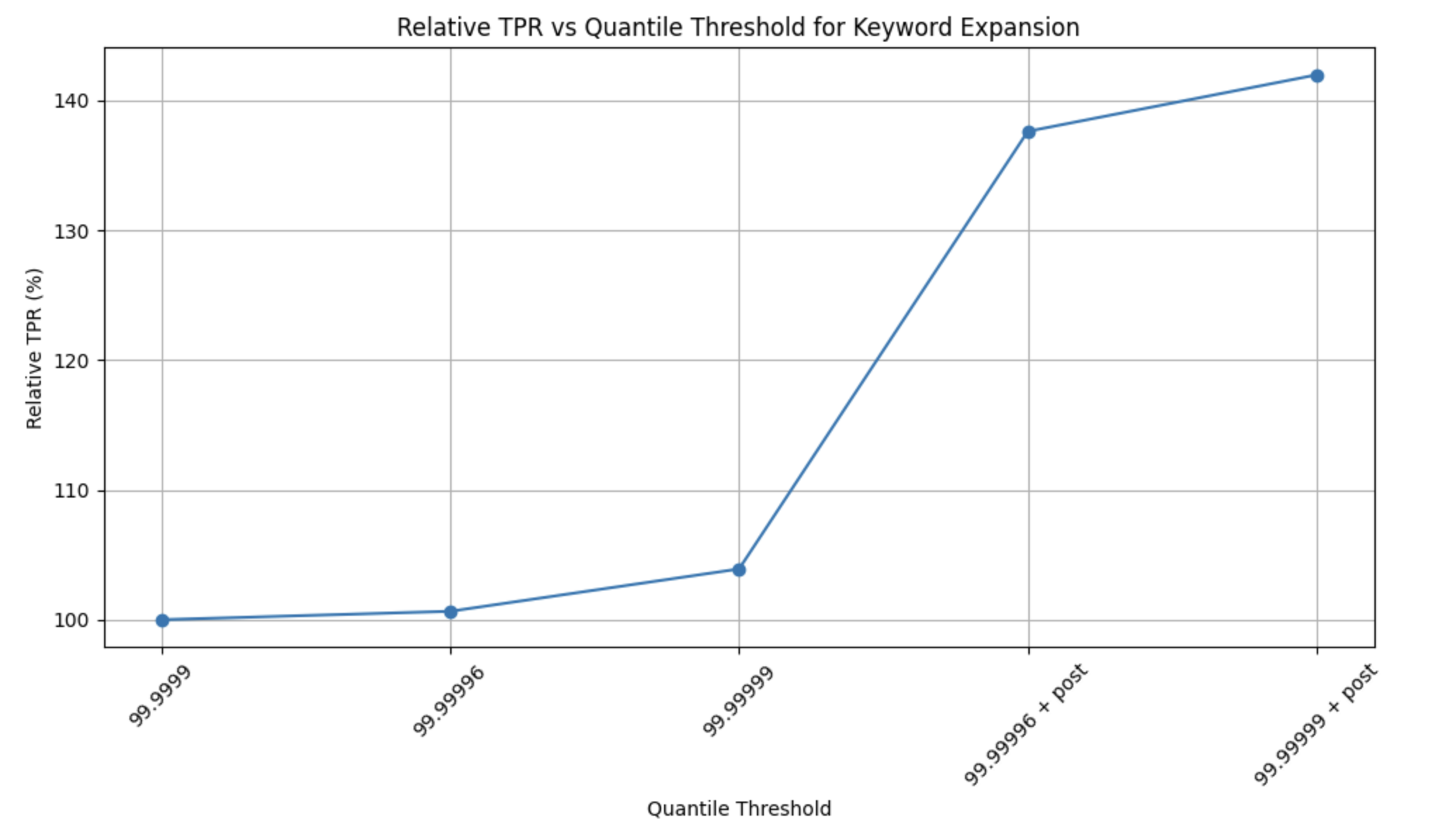}
    \caption{TPR vs Similarity Quantile thresholds. Lower quantiles introduce many semantically unrelated expansions. Post-processing (e.g., gender and numeric consistency filters) significantly improves TPR, especially at higher thresholds. TPR at the 99.9999th quantile is normalized to 100\%.}
    \label{fig:quantile_tpr}
  \end{subfigure}
  \hfill
  \begin{subfigure}[b]{0.49\linewidth}
    \centering
    \includegraphics[width=\linewidth]{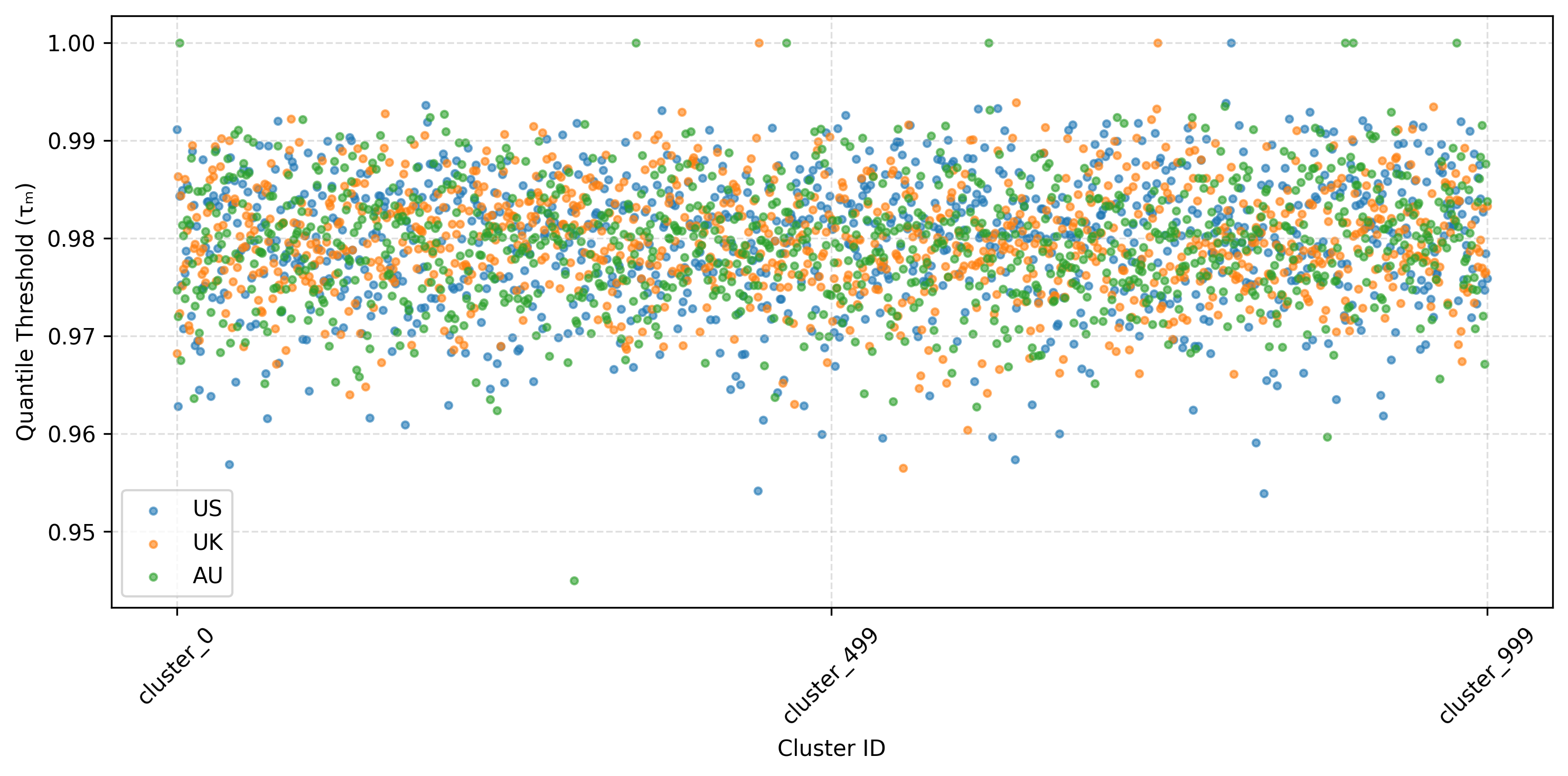}
    \caption{Quantile threshold ($\tau_m$) for each cluster $C_m$. Average threshold centers around $0.98$, but variance indicates heterogeneous cluster compactness and distance distributions, motivating adaptive per-cluster thresholds.}
    \label{fig:threshold_distribution}
  \end{subfigure}
  \caption{(a) TPR trends across quantile thresholds, and (b) distribution of quantile thresholds across clusters.}
  \label{fig:side_by_side}
\end{figure}


\begin{table}[t]
\centering
\caption{Document Side expansion of Original Keyword}
\resizebox{\columnwidth}{!}{%
\begin{tabular}{l|l}
\toprule
\textbf{Original Keyword} & \textbf{Semantic Expansions} \\
\midrule
LED Garden Lights & Outdoor LED Lights, Garden Lighting, Landscape Lights, Solar Garden Light \\
Ladies Winter Jumpers &  Women's Winter Sweaters, Knitted Jumpers for Women, Cozy Winter Tops \\
Running Shoes White & White Athletic Shoes, White Sneakers, Sports Footwear, Running Sneakers \\
65W USB-C GaN Charger & USB-C GaN Power Adapter 65W, GaN Technology Charger 65W, 45W, 30W, USB-C GaN Adapter 65W \\
\bottomrule
\end{tabular}%
}
\label{tab:Expansions}
\end{table}

\textit{Cluster threshold selection} To determine appropriate similarity thresholds for filtering expanded keywords, we conducted a systematic evaluation of quantile-based thresholds at the cluster level. We sampled 100 randomly selected AdKeywords along with their nearest neighbors retrieved from the embedding space. For each keyword, we evaluated expansions generated at several similarity quantile cutoffs $p$: 95\%, 99\%, 99.99\%, 99.999\%, 99.9999\%, 99.99996\%, and 99.99999\%. At lower quantiles (e.g., 95\%--99.999\%), we observed a substantial number of semantically unrelated or mismatched expansions. These included incorrect gender substitutions (e.g., ``men’s shoes'' expanded to ``women’s sandals'') and numerical inconsistencies (e.g., ``iPhone 13 case'' matching ``iPhone 12 accessories''), which are particularly problematic in the e-commerce domain where attribute precision is critical. To mitigate these issues, we introduced a lightweight post-processing pipeline that filters expanded keywords for gender and numeric consistency with the original term. Evaluation, assisted by ChatGPT-3.5, was used to judge the semantic validity of the expansions at each quantile threshold. We measured True Positive Rate (TPR) both before and after post-processing and found that TPR significantly improves at higher quantiles (e.g., $\geq 99.9999\%$), especially after applying filters. In addition, we analyzed the distribution of per-cluster thresholds ($\tau_m$), computed as the $p$-th quantile of intra-cluster similarity distances. While the global average threshold centers around 0.98, we observed high variance across clusters, indicating heterogeneity in cluster compactness. This reinforces the need for adaptive, cluster-specific thresholds rather than a fixed global cutoff.Notably, we also found a positive correlation between cluster size and similarity threshold. Larger clusters—typically corresponding to broader, more ambiguous e-commerce concepts like ``case'' or ``perfect shoes''—exhibited higher threshold values, reflecting the semantic spread of such concepts. In contrast, smaller, tightly packed clusters (e.g., niche or highly specific product types like ``65W USB-C GaN charger'') received lower thresholds, ensuring only precise matches are retained. This observation highlights the importance of adjusting threshold strictness based on cluster-level density to balance recall for general terms and precision for specific ones. Based on this analysis, we selected the optimal quantile value ($p=99.9999$) as our production baseline. This value provides strong recall for common concepts while maintaining high-quality semantic expansion after post-processing. Figure~\ref{fig:quantile_tpr} illustrates the relationship between quantile thresholds and TPR performance, and Figure~\ref{fig:threshold_distribution} shows the per-cluster threshold distribution and its correlation with cluster size. Table~\ref{tab:Expansions} shows some expansions attached to the original keyword. The expansions are matched to the query, broadening the reach of the original keywords and the sellers' ad groups. Once the sellers' items pass the downstream filters, they are displayed on the final page, enhancing both their visibility and overall reach.


\subsubsection{Offline Evaluation of Relevance Adjustment:} 
Our relevance model outputs a regression score on a scale of 0 to 5, aligned with human judgment labels (Perfect = 5, Excellent = 4, Good = 3, Fair = 2, Bad = 1). Offline evaluation of the stacked model on a held-out dataset,which had items from keyword expansion inventory, was conducted. It was found that compared to the production relevance model, it reduces RMSE for items that were annotated by human as Fair/Bad by >4\% and items that were labeled Excellent by > 1\% respectively. 
Table ~\ref{tab:query_items} presents some example query-items from keyword expansion and how their relevance score is changed by our stacked model.

\begin{table}[t]
\centering
\caption{Example query-items from keyword expansion framework and their predicted relevance score comparison. We have redacted the identifiers in the examples for privacy. ↑ and ↓ are added to the Adj. Rel column to demonstrate change in relevance by the Stacked Model.}
\label{tab:query_items}
\resizebox{\columnwidth}{!}{%
\begin{tabular}{p{3.5cm} p{6.3cm} p{1.5cm} p{1.5cm} p{1.5cm}}
\toprule
\textbf{Query} & \textbf{Item Title} & \textbf{Old Rel.} & \textbf{Adj. Rel} & \textbf{True Rel} \\
\midrule
tory burch camera & SmallRig \censor{4}-Arm Suction Cup Camera Mount Kit, \censor{15 kg} Load capacity \censor{SC-15K -3565} & 1.086 & 0.839 ↓ & Bad \\
4.4mm balanced cable & Blood Sugar Blend - Balance Sugar Glucose Levels - \censor{2} Pack & 1.022 & 0.776 ↓ & Bad \\
slouchy bag & Genuine Leather Purses and Handbags Women Vintage Shoulder Bag Crossbody \censor{Bag} & 2.532 & 2.606 ↑ & Fair \\
anker power bank & Power Bank Charging Portable External Battery Backup For Phone Laptops Tablet & 2.724 & 2.793 ↑ & Good \\
humidistat dehumidifier & ANDTE \censor{250} Pints Commercial Dehumidifier for Basement with Drain Pump & 2.775 & 2.901 ↑ & Good \\
\bottomrule
\end{tabular}%
}
\end{table}

\begin{table}[t]
\centering
\caption{Percentage lift from the baseline in A/B test performance metrics.}
\resizebox{\columnwidth}{!}{%
\begin{tabular}{lcccccc}
\toprule
Method & CTR & Impression & BI/Click & Slot Neutral CTR & CPC & Avg Relevance \\
\midrule
Emb only & -3.1 & 7.48 & -3.12 & -1.23 & 4.72 & -0.47 \\
Emb+Cluster threshold & -1.26 & 3.74 & -1.53 & -0.65 & 3.52 & -0.25 \\
Emb+Cluster threshold+Relevance model & -0.62 & 3 & -1.96 & -0.13 & 3.05 & -0.18 \\
\bottomrule
\label{tab:ab_results}
\end{tabular}
}
\end{table}

\subsection{Business Metrics}
We conducted a two-week online A/B experiment to evaluate the effectiveness of our proposed enhancements to Broad Match, where a buyer’s query is matched to an expanded keyword if the expanded keyword is a subset of the query tokens. This containment-based matching strategy offers a balance between \textit{semantic flexibility} and \textit{interpretability}, reducing ambiguity while expanding reach.The experiment tested three system variants:
\begin{enumerate}
    \item \textbf{Embedding-only keyword expansion}
    \item \textbf{Embedding with cluster-based thresholding}
    \item \textbf{Embedding with both cluster-based thresholding and relevance model refinement}
\end{enumerate}

Each variant was compared against a baseline system that uses token-based expansion—where expansion is limited to simple lexical variations (e.g., \textit{iPhone} $\rightarrow$ \textit{iPhones}, \textit{iPhone case}). This baseline has minimal brand conquesting and limited semantic generalization.To evaluate impact, we used normalized system-level metrics including: \textit{Impressions (Imps)}, \textit{Ad Revenue (Rev)}, \textit{Click-Through Rate (CTR)}, \textit{Cost-Per-Click (CPC)}, and \textit{Bought Items per Click (BI/click)}. We report relative percentage improvements over a shared control baseline to preserve privacy while capturing directional performance shifts.
As shown in Table~\ref{tab:ab_results}, introducing cluster-based thresholding followed by relevance model refinement leads to progressive improvements across most metrics. While the initial embedding-only variant increased impressions but reduced CTR, subsequent refinements significantly recovered CTR and yielded net gains in BI/click and revenue. This demonstrates that precision tuning and relevance filtering are key to high-quality expansion. Importantly, the expanded system enabled broader keyword targeting, improving ad coverage and inventory utilization. Though there was a slight dip in CTR, \textit{slot-normalized CTR} remained neutral, and the expanded queries matched to listings with higher \textit{average selling price (ASP)}—leading to higher \textit{gross merchandise bought (GMB)}. Furthermore, as competition increased in these newly matched segments, \textit{average bid levels} rose, increasing \textit{clearing prices} and \textit{Cost-Per-Click (CPC)}. While this impacted \textit{Return on Ad Spend (ROAS)} for some sellers, it also reflected increased visibility and competitiveness, particularly in high-value verticals.
Overall, the enhanced system delivered a net-positive impact on total ad revenue, provided sellers with greater velocity, and demonstrated strong potential for scaling intelligent keyword expansion with minimal manual effort.

\section{Conclusion}
Traditional keyword-based matching systems remain fundamental in online advertising but struggle with semantic variability and long-tail queries. We introduced a semantic keyword expansion framework that enriches advertiser-provided keywords on the document side using dense embeddings, scalable nearest neighbor retrieval, and adaptive cluster-based thresholding. This method enhances recall and precision without altering existing token-based infrastructures. Future work includes incorporating query-side signals (e.g., click feedback), multilingual extensions, real-time adaptive thresholds based on user interactions, and personalized keyword expansion to further boost scalability, adaptability, and relevance in dynamic e-commerce settings.


\bibliography{reference}

\end{document}